\documentclass[runningheads]{llncs}
\usepackage[T1]{fontenc}
\usepackage{graphicx}
\usepackage{pifont}    
\newcommand{\cmark}{\ding{51}}  
\newcommand{\xmark}{\ding{55}}  
\usepackage{array}
\usepackage{subcaption}

\usepackage{xcolor}
\usepackage{tabularx}
\usepackage{threeparttable}
\usepackage{tablefootnote}
\newcolumntype{C}[1]{>{\centering\arraybackslash}m{#1}}
\usepackage{comment}
\usepackage{amsmath}
\usepackage{booktabs}

\begin{document}
\title{From Legacy to Standard: LLM-Assisted Transformation of Cybersecurity Playbooks into CACAO Format}
\titlerunning{LLM-Assisted Transformation of Cybersecurity Playbooks}
%
\author{Mehdi Akbari Gurabi\inst{1,2}\orcidID{0000-0002-1734-8367} \and
Lasse Nitz \inst{1,2}\orcidID{0000-0002-3131-7444} \and
Radu-Mihai Castravet \inst{2}\orcidID{} \and Roman Matzutt \inst{1}\orcidID{0000-0002-4263-5317} \and Avikarsha Mandal \inst{1}\orcidID{0000-0002-8641-7207} \and Stefan Decker \inst{1,2}\orcidID{0000-0001-6324-7164}
}
\authorrunning{M. Akbari Gurabi et al.}
%
\institute{Fraunhofer FIT, Sankt Augustin, Germany
\email{\{mehdi.akbari.gurabi,lasse.nitz,roman.matzutt,avikarsha.mandal, stefan.decker\}@fit.fraunhofer.de} \and
RWTH Aachen University, Aachen, Germany\\
\email{\{mehdi.akbari.gurabi,radu.castravet\}@rwth-aachen.de}}
\maketitle              

\begin{abstract}
Existing cybersecurity playbooks are often written in heterogeneous, non-machine-readable formats, which limits their automation and interoperability across Security Orchestration, Automation, and Response platforms.
This paper explores the suitability of Large Language Models, combined with Prompt Engineering, to automatically translate legacy incident response playbooks into the standardized, machine-readable CACAO format. We systematically examine various Prompt Engineering techniques and carefully design prompts aimed at maximizing syntactic accuracy and semantic fidelity for control flow preservation. Our modular transformation pipeline integrates a syntax checker to ensure syntactic correctness and features an iterative refinement mechanism that progressively reduces syntactic errors. We evaluate the proposed approach on a custom-generated dataset comprising diverse legacy playbooks paired with manually created CACAO references. The results demonstrate that our method significantly improves the accuracy of playbook transformation over baseline models, effectively captures complex workflow structures, and substantially reduces errors. It highlights the potential for practical deployment in automated cybersecurity playbook transformation tasks.


\end{abstract}

\section{Introduction}

Cybersecurity playbooks are predefined workflows that outline step-by-step procedures, including detection, containment, eradication, recovery, and post-incident review for specific cybersecurity incidents \cite{incident_response_ibm}. They provide Security Operations Center (SOC) teams with structured and consistent guidance, ensuring rapid decision making, clear assignment of roles, and adherence to organizational policies during critical security events. When an incident is detected, SOC teams use these playbooks to efficiently initiate response actions and contain threats promptly. Following containment, the playbooks guide SOC teams through the recovery and post-incident phase, ensuring thorough documentation of all activities, which is essential for regulatory compliance. For example, the General Data Protection Regulation (GDPR) requires detailed records of responses to personal data breaches. Furthermore, EU directives such as NIS2 emphasize the importance of cross-border collaboration and standardized incident response, advocating predefined and clearly documented procedures, such as playbooks, to facilitate effective, transparent, and accountable cybersecurity management \cite{SAPPAN}. Similarly, standards such as ISO/IEC 27001 require comprehensive documentation of security incidents and recovery measures. Consequently, standardized playbooks enhance accountability, ensure traceability, and support compliance with these regulatory frameworks.

However, despite their critical role, many existing cybersecurity playbooks are manually maintained in unstructured or semi-structured formats, rendering them non-machine-readable and non-executable. The legacy unstructured playbooks are typically written in narrative, natural language (i.e., free text) and often exist in formats such as PDFs, Word documents, internal wikis, emails, or operational manuals. This significantly hinders interoperability, automation, and the effective sharing and reuse of best practices across organizational boundaries \cite{akbari2022sasp}. This fragmentation limits the efficiency and speed of collaborative incident response, ultimately impacting overall cybersecurity resilience \cite{ERCIM139}. To address these interoperability challenges, the Collaborative Automated Course of Action Operations (CACAO) standard was developed, defining a vendor-agnostic JSON schema \cite{cacaov2_spec}. CACAO playbooks enable seamless exchange across organizations without requiring rewriting, and can be automatically executed by Security Orchestration, Automation and Response (SOAR) platforms, significantly enhancing incident response speed and effectiveness \cite{cybersecurity_playbook_sharing_with_stix}. 
However, transitioning legacy playbooks to the CACAO format involves a significant manual transformation effort. Mapping nested workflows, control flow branches, and action definitions into the structured CACAO schema is both time consuming and error-prone.

This paper addresses these challenges by proposing and evaluating a systematic solution that translates various legacy playbook formats into a standardized, machine-readable representation aligned with the CACAO specification with the help of Large Language Models (LLMs). LLMs, pre-trained on massive text corpora, including cybersecurity manuals, have shown advancements in machine translation, entity extraction, and structured data generation \cite{llms_are_few_shot_learners,template_based_named_entity_recognition}. However, out-of-the-box LLM outputs can hallucinate or violate schema constraints. Enhancement techniques such as prompt engineering aim to address these issues. We systematically examine various Prompt Engineering techniques, identify the most effective methods, and carefully design natural language prompts aimed at improving both syntactic accuracy and semantic fidelity in the transformation process. We define the transformation process as the conversion of legacy playbooks into valid CACAO JSON, and we investigate whether prompt engineered LLMs can produce syntactically valid outputs, preserve semantic fidelity (control flow, actions, variables) and be refined via iterative feedback to eliminate residual errors. Due to the absence of publicly available datasets, we generate a custom evaluation dataset consisting of community playbooks across multiple formats paired with their corresponding manually translated CACAO versions, enabling benchmarking of state-of-the-art LLMs.

\textbf{Contributions:}  This paper makes four primary contributions: (a) a prompt engineering taxonomy tailored specifically for CACAO transformation, covering techniques such as Persona, Template, Chain-of-Thought, Direct Knowledge Injection, and Few-Shot, while highlighting the trade-offs between cost and accuracy; (b) a modular pipeline that integrates task decomposition, structured prompt assembly, a CACAO syntax checker, and an iterative feedback loop to progressively correct errors; (c) a custom evaluation dataset composed of structured SOAR community playbooks \cite{awesome_playbooks} along with their corresponding manually translated CACAO versions as ground truths; and (d) an extensive evaluation of state-of-the-art LLMs using syntactic error counts, and semantic similarity metrics including Damerau–Levenshtein distance and Graph Edit Distance\footnote{The source code, prompts, dataset, and evaluation results are available at \url{https://github.com/Fraunhofer-FIT-DSAI/CyberGuard}.}.

\section{Background and Related Work}
\label{sec:background}

\begin{figure}[ht]
\centering
\includegraphics[width=0.5\linewidth]{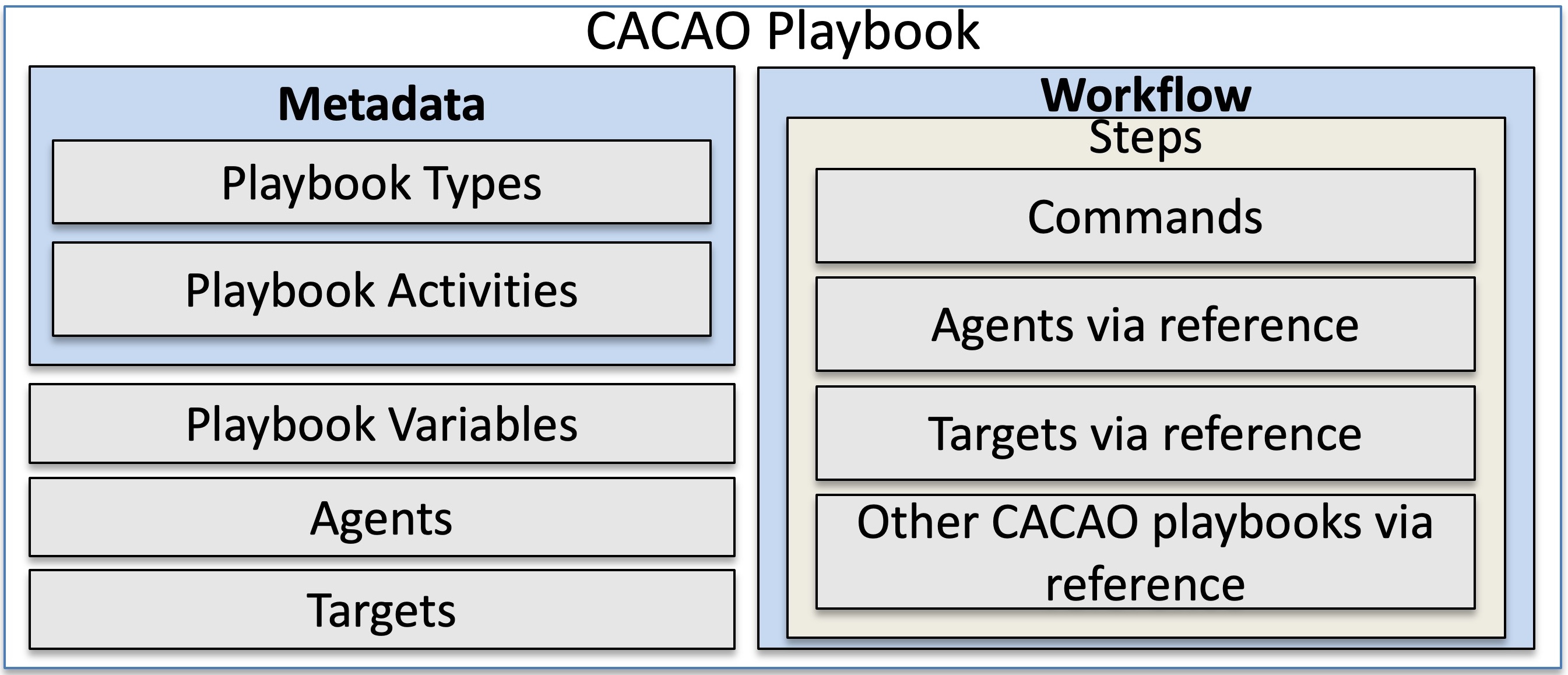}	\caption{Simplified CACAO playbook structure based on \cite{cacaov2_spec}.}
\label{fig:cacao}
\end{figure}

\textbf{Cybersecurity Playbooks:} Modern incident response relies on well-defined cybersecurity playbooks, semi-automated workflows that guide security teams \cite{req}. Traditionally, these playbooks are expressed in ad hoc formats (documents, spreadsheets, bespoke JSON), making them laborious to share, adapt, or execute automatically. To address this, the CACAO specification provides a vendor-agnostic, structured, and standardized JSON schema for cybersecurity playbooks \cite{cacaov2_spec}. The current version of the CACAO specification is 2.0, which includes resource updates from the previous version. CACAO playbooks define workflows that comprise logically ordered steps that enable organizations to systematically detect, investigate, prevent, mitigate, and remediate cybersecurity threats effectively. The CACAO structure encompasses key properties such as metadata, workflow definitions, playbook variables, agents, and targets, as shown in Figure \ref{fig:cacao}). Important metadata fields include \textit{$playbook\_types$}, which describe operational roles (e.g. detection), and \textit{$playbook\_activities$}, specifying detailed actions such as \textit{scan-system} for a potentially compromised device. Central to CACAO is the richly expressive \texttt{workflow} model with eight step types (start, end, action, conditional, loop, parallel, switch, playbook‐action). It organizes actions into sequences or parallel execution flows and supports advanced logic constructs such as conditions, loops, and nested playbook invocations, which enables complex cybersecurity processes.

The CACAO standard provides a vendor-agnostic schema for playbooks, unifying metadata (e.g., \texttt{playbook\_types}, \texttt{playbook\_activities}), variables, agents, targets, and a richly expressive \texttt{workflow} model with eight step types (start, end, action, conditional, loop, parallel, switch, playbook‐action) \cite{cacaov2_spec}. In addition, recent work by Tsirakis et al. \cite{tsirakis2025} demonstrates ongoing advancements in CACAO-centric developments, presenting a Knowledge Management System for the lifecycle management of CACAO playbooks. \\

\textbf{LLMs and Prompt Engineering:} LLMs have revolutionized natural language processing (NLP) by taking advantage of the self-attention mechanism of the transformer architecture for highly parallelizable pretraining in massive corpora \cite{attention_is_all_you_need,comprehensive_overview_of_llms}. Early pipelines required expensive fine-tuning for each downstream task, but new models demonstrated that a single, task-agnostic LLM can perform new tasks with carefully crafted prompts and a few in-context examples \cite{llms_are_few_shot_learners}. Key advances include expanding context windows to maintain coherence over long inputs \cite{extending_context_window}, and instruction-tuning to align models with human instructions without task-specific weight updates \cite{finetuned_llms_are_zero_shot_learners}. Prompt engineering guides LLM output without changing their parameters \cite{systematic_survey_prompt_engineering_in_llms}. Examples of common patterns are listed in Table \ref{tab:example}.

\begin{table}[ht]
\centering
\scriptsize
\caption{Example of prompt engineering techniques.}
\begin{tabular}{p{2.15cm}|p{10.1cm}}
\hline
\textbf{Technique} & \textbf{Description} \\\hline
Persona & “Act as an expert in a field” to bias domain-relevant language \cite{quantifying_persona_effect}. \\ \hline
Template & Enforce a specific schema output via placeholders \cite{prompt_pattern_catalog}. \\\hline
Reflection & Request explanations of reasoning to surface assumptions \cite{prompt_pattern_catalog}. \\\hline
Chain-of-Thought (CoT) & Elicit intermediate reasoning steps for complex tasks \cite{chain_of_thought}, with a zero-shot variant 'Let's think step by step' \cite{zero_shot_chain_of_thought}. \\\hline
Least-to-Most & Decompose problems into subproblems and solve incrementally \cite{least_to_most_prompting}. \\\hline
Retrieval-Augmented Generation (RAG) & Fetch external context to supplement model knowledge \cite{fine_tuning_or_retrieval}, although naive RAG can introduce irrelevant or hallucinated content \cite{rag_survey}. \\\hline
Direct Knowledge Injection & Embed precise snippets in the prompt to constrain outputs without external retrieval. \\\hline
Few-Shot Prompting & Provide a handful of input–output examples in the prompt, markedly boosting performance \cite{llms_are_few_shot_learners}; trades higher token usage against zero-shot prompting, which relies only on task descriptions \cite{systematic_survey_prompt_engineering_in_llms}. \\ \hline
\end{tabular}
\label{tab:example}
\end{table}

Together, the structured schema of CACAO and advanced prompt engineering patterns create a promising pathway to automate the transformation of existing playbooks into the CACAO standard.

\textbf{Related Work:} Prior efforts have demonstrated the power of prompt engineered LLMs to extract structured information and generate formal process models from natural language. Vijayan \cite{prompt_engineering_approach_structured_data_extraction} used Google Bard and ChatGPT to pull tuples (e.g., traveler name, origin, dates) from unstructured email text, combining Persona, Few-Shot and CoT prompts.  They report that Few-Shot prompting yielded the best precision–recall trade-off, while CoT improved reasoning over complex extraction tasks. Polak and Morgan’s ChatExtract pipeline \cite{extracting_accurate_materials_data} also uses LLM in a three-stage approach: identify relevant passages, extract triplets of unit value of material and verify by follow-up queries to mitigate hallucinations. However, their focus remains narrow (single-triplet schema) compared to the multilayered, nested structure of CACAO playbooks.

Licardo et al. \cite{method_extracting_bpmn} and Kourani et al. \cite{process_modeling_with_llms} pushed LLMs into business process modeling, from free-text descriptions they generate BPMN diagrams, adopting Role-Prompting (similar to Persona), Knowledge Injection, and Few-Shot techniques.  Kourani et al. also embed an interactive feedback loop, allowing users to correct model missteps.  Their evaluations show the strength of GPT-4 in capturing non-hierarchical dependencies \cite{gpt4_technical_report}. However, these methods are insufficient for translating cybersecurity playbooks because the structured outputs they handle are typically less complex than cybersecurity playbooks. Thus, directly applying BPMN-oriented methods does not adequately address the semantic richness required for cybersecurity incident response. Our approach extends beyond these foundational studies in two significant ways.
Firstly, translating legacy cybersecurity playbooks demands handling deeply nested workflow logic, conditional branches, parallel, and loop constructs, far more complex than flat tuple extraction or BPMN generation. Secondly, we systematically compare multiple prompt engineering patterns (Persona, Template, CoT, Knowledge Injection, Few-Shot) to optimize both syntactic validity and semantic fidelity of the resulting CACAO JSON, rather than focusing on a single technique or schema. This comprehensive evaluation enables us to identify the most effective approaches for capturing the nuanced structure of cybersecurity playbooks.


\section{Overview of LLM-Assisted Playbook Transformation Framework}
\label{sec:methodology}

\textbf{Requirements:} Transformation success is governed by six primary requirements listed in Table~\ref{tab:requirements}. In particular, the translation process must meet several critical requirements to ensure effective adoption. These include producing machine-readable CACAO JSON output, syntactic accuracy which means ensuring strict compliance with the CACAO JSON schema, semantic fidelity for maintaining the original workflow logic and metadata integrity, cost-efficiency by optimizing LLM token usage, adherence to LLM context-window constraints, and optimizing the implementation effort. Balancing these requirements is essential to create a robust, scalable, and practical transformation pipeline.
        
\begin{table}[ht]
\centering
\scriptsize
\caption{Key requirements for automated playbook transformation.}
\label{tab:requirements}
\begin{tabular}{p{0.27\linewidth}|p{0.72\linewidth}}
\toprule
\textbf{Requirement}      & \textbf{Description}                                                                                 \\
\midrule
Ensuring formatted output               & Outputs must be emitted as valid CACAO JSON.          \\
Syntactic accuracy        & Generated playbooks must conform to CACAO 2.0 specification without manual fixes.            \\
Semantic fidelity         & control flow constructs, conditional branches, and action definitions must match the source playbook’s intended logic.     \\
Token budget              & Prompt and response token counts should be minimized to control API usage costs.                     \\
Staying within Context window/output token limit          & All individual sub‐task prompts must stay within the model’s maximum token capacity (input and output) to avoid truncation.         \\
Moderate implementation effort     & The transformation pipeline should balance performance improvements against engineering complexity and facilitate adoption.   \\
\bottomrule
\end{tabular}
\end{table}

\textbf{System Architecture:} Our approach decomposes the complex task of translating heterogeneous playbook formats into the CACAO schema into a modular pipeline, combining a Prompt Engineering Module (PEM), a schema‐based Syntax Checker, and an optional human–in–the–loop feedback loop.
This design ensures that each stage remains focused on a single responsibility: prompt generation, syntactic validation, or error remediation. The overall orchestration handles large playbooks without exceeding model context or output limits.

\begin{figure}[ht]
  \centering
  \includegraphics[width=0.9\linewidth]{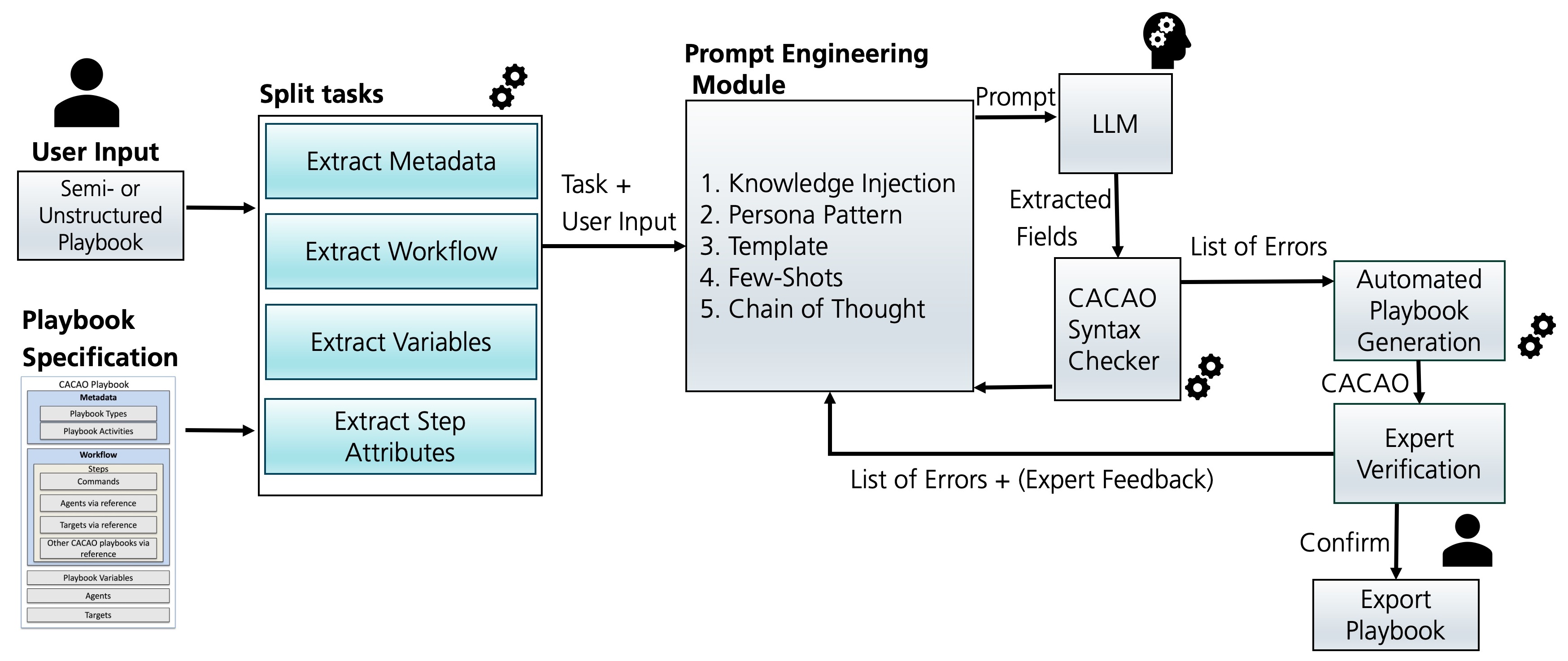}
  \caption{Modular transformation pipeline from decomposition of task, to PEM generates focused prompts, LLM produces JSON fragments, Syntax Checker validates against CACAO schemas, and Feedback Loop refines invalid fragments.}
  \label{fig:architecture}
\end{figure}

Figure~\ref{fig:architecture} depicts the high‐level architecture of our transformation pipeline.  First, the PEM constructs a series of sub‐task prompts, each targeting a specific CACAO entity (metadata, workflow skeleton, step attributes, variables).  These prompts leverage selected engineering patterns to guide the LLM toward valid and semantically sound outputs. The selected engineering prompts are Persona, Template, CoT, Direct Knowledge Injection, and Few‐Shot. Each JSON fragment produced by the LLM is then passed to the Syntax Checker, which enforces consistency against a patched set of CACAO 2.0 JSON schemas, immediately flagging omissions or type mismatches.  Any fragments that fail validation would be routed into the feedback loop, where an analyst optionally reviews error logs, issues corrections via natural language, and those corrections in addition to the automatically generated syntax errors are fed back into the PEM for iterative refinement.

\textbf{Task Decomposition:} Rather than prompting the LLM to translate an entire playbook at once, a strategy prone to context overload and cascading errors, we manually decompose the transformation into four sequential subtasks.  First, metadata extraction retrieves top‐level fields such as \texttt{name}, \texttt{description}, and \texttt{playbook\_types}.  Next, the workflow skeleton step identifies all playbook steps by name, assigns unique identifiers, and classifies each into one of the eight CACAO step types.  The third subtask, step attributes extraction, iterates over each skeleton step to pull type-specific properties (e.g. \texttt{on\_success}, \texttt{commands}, conditional expressions), injecting the appropriate CACAO schema snippet to constrain the model.  Finally, variable extraction collates both global and per‐step variables, capturing names, types, and descriptions. The prompt for each subtask includes only its immediate input plus a minimal history, ensuring that the model stays within its context window and focuses on a narrowly defined goal.  This decomposition not only reduces token usage and error rates, but also simplifies error isolation and iterative refinement.  

\textbf{Prompt Engineering Taxonomy:} To systematically explore how different prompt styles affect transformation quality, we categorize core patterns and analyze their relationships with the requirements in Table \ref{table:approaches-requirements}.  \emph{Persona} prompts (“Act as a cybersecurity playbook translator”) inject domain context at negligible cost and yield small improvements in both syntax and meaning.  The \emph{Template} pattern enforces strict JSON output at zero additional cost, significantly reducing schema violations.  \emph{Reasoning} combines zero‐shot Chain‐of‐Thought cues (“Let’s think step‐by‐step”) with a Reflection request (“Justify your answer”), adding roughly an overhead token 8\% and producing mixed gains, useful for complex conditional logic but sometimes verbose.  \emph{Knowledge Injection} embeds brief excerpts from the CACAO specification, incurring 5–10\% more tokens but delivering large reductions in syntax errors and moderate semantic gains.  Finally, a single example \emph{One‐Shot} illustrates the target JSON format at a cost of 5–20\% more tokens and yields a medium improvement in syntactic validity and small semantic benefits. An overview of the potential approaches and their relationships with the requirements is illustrated in Table \ref{table:approaches-requirements}.

\begin{table}[h]
	\centering
        \scriptsize
	\caption{Potential approaches against requirements. A checkmark (\cmark) indicates that the technique either fulfills or is unlikely to violate the specified requirement. A cross (\xmark) suggests that the technique may violate the requirement, while a dash (-) indicates a neutral or unknown impact. }
	\label{table:approaches-requirements}
	\begin{tabular}{C{1.9cm}|C{1.4cm}|C{1.4cm}|C{1.4cm}|C{2.0cm}|C{1.4cm}|C{1.8cm}}
		\textbf{Approach / Requirement}           & Maximizing  Syntactic  Accuracy & Maximizing  Semantic  Accuracy & Minimizing  Cost &  Staying  Within the Model’s Maximum Token Capacity  & Ensuring  a Formatted  Output & Minimizing  Implementation  Complexity \\
		\hline
		Persona  Pattern                          & -                               & \cmark                         & \cmark           & \cmark                               & -                             & \cmark                                 \\
		\hline
		Template  Pattern                         & -                               & -                              & \cmark           & -                                      & \cmark                        & \cmark                                 \\
		\hline
		Few-Shot                                  & \cmark                          & \cmark                         & \xmark                      & \cmark                                 & -                             & \xmark                                 \\
		\hline
		CoT                                       & \cmark                          & \cmark                         & \xmark                    & \xmark                                 & -                             & \xmark                                 \\
		\hline
		Zero-Shot  CoT  with  Reflection  Pattern & -                               & \cmark                         & \cmark           & \xmark                                 & -                             & \cmark                                 \\
		\hline
		Manual  Decomposition                     & \cmark                          & \cmark                         & \xmark                   & \cmark                                 & -                             & \xmark                                 \\
		\hline
		Direct  Knowledge  Injection              & \cmark                          & \cmark                         & \xmark                    & \xmark                                 & -                             & \cmark                                 \\
		\hline
		Fine-tuning                               & \cmark                          & \cmark                         & \xmark           & -                                                   & \cmark                        & \xmark                                 \\
	\end{tabular}
	\label{tab:approach-requirements}
\end{table}

\section{Implementation}
\label{sec:implementation}

Our transformation pipeline is implemented in Python, leveraging three state‐of‐the‐art LLMs to explore trade‐offs between cost, context capacity, and fidelity. We first compared different language models in Table \ref{table:llm-comparison} in three categories of ultra-large, mid-sized, and compact models. Then we selected one from each category for the experiments. We evaluate GPT-4o and its lower-cost and smaller sibling GPT-4o-mini, as well as the open-source Llama3.1-8B compact model, each with up to 128K token context windows. Prompt orchestration and caching are handled via LangChain’s prompt‐assembly framework, with an SQLite back-end to avoid redundant API calls during development. 

Throughout the implementation phase, we aimed to ensure reproducibility of our approach's results. However, due to the probabilistic nature of the Transformer architecture behind LLMs \cite{attention_is_all_you_need}, this was not entirely possible. Despite the stochastic output of LLMs \cite{llms_are_few_shot_learners}, we partially limited result fluctuations by setting the \emph{temperature} parameter to zero. This parameter regulates randomness, leading to more diverse outputs as it increases \cite{temperature_creativity}. We achieved reproducible results only for LLama3.1. For OpenAI models, even with zero temperature, we observed variability in tasks like extracting the \textit{on\_completion} step parameter. To mitigate this, we used the \emph{seed} parameter as recommended in the OpenAI documentation\footnote{\url{https://platform.openai.com/docs/advanced-usage/reproducible-outputs} - Accessed 14.09.2024}, setting it to 42, but it did not fully support reproducibility. Likely, the seed parameter is supported only by specific models like GPT-3.5-Turbo\footnote{\url{https://cookbook.openai.com/examples/reproducible\_outputs\_with\_the\_seed\_parameter} - Accessed 14.09.2024}.

\begin{table}[h]
    \centering
    \begin{threeparttable}
    \caption{Comparison of LLMs: The models are clustered according to their price \protect\footnotemark[6]. The data comes from the Artificial Analysis LLM leaderboard\protect\footnotemark[7] unless specified otherwise.}
        \label{table:llm-comparison}
        \scriptsize
        \begin{tabular}{|p{5.5cm}|l|l|l|l|l|}
            \hline
            \textbf{Model Name} & \textbf{Context} & \textbf{Price}\protect\footnotemark[6] & \textbf{Type} & \textbf{MMLU (\%)} & \textbf{DROP (f1)}   \\ \hline
 GPT-4o              & 128k             & 4.38\$                         & closed        & 88.7               & 83.4\protect\footnotemark[8] \\ \hline
 Gemini 1.5 Pro      & 1M               & 5.25\$                         & closed        & 85.9               & 78.9\protect\footnotemark[8] \\ \hline
 Claude 3 Opus       & 200k             & 30\$                           & closed        & 86.8               & 83.1\protect\footnotemark[8] \\ \hline \hline
 GPT-4o-mini         & 128k             & 0.26\$                         & closed        & 82                 & 79.7\protect\footnotemark[9] \\ \hline
 GPT-3.5-turbo       & 16k              & 0.75\$                         & closed        & 70                 & 70.2\protect\footnotemark[9] \\ \hline
 Gemini 1.5 Flash    & 128k             & 0.13\$                         & closed        & 78.9               & 78.4\protect\footnotemark[9] \\ \hline
 Claude 3 Haiku      & 128k             & 0.5\$                          & closed        & 75.2               & 78.4\protect\footnotemark[9] \\ \hline \hline
 Llama3.1 8B         & 128k             & 0\protect\tnote{*}                     & open          & 66.7\protect\tnote{\S}     & 59.5\protect\tnote{\S}       \\ \hline
 Mistral 7B          & 8k               & 0\protect\tnote{*}                     & open          & 63.6\protect\tnote{\S}     & 53\protect\tnote{\S}         \\ \hline
 Gemma 7B            & 8k               & 0.07\$                         & closed        & 64.3\protect\tnote{\S}     & 56.3\protect\tnote{\S}       \\ \hline
 
        \end{tabular}

        \begin{tablenotes}
            \item[*] denotes that the model can be run locally and therefore no "token cost" is incurred
            \item[\S] denotes the fact that the data comes from \cite{llama3}
        \end{tablenotes}
    \end{threeparttable}
\end{table}

\footnotetext[6]{ Price definition from \url{https://artificialanalysis.ai/methodology} - Accessed 09.05.2025}
\footnotetext[7]{ \url{https://artificialanalysis.ai/leaderboards/models} - Accessed 09.05.2025}
\footnotetext[8]{ \url{https://openai.com/index/hello-gpt-4o/} - Accessed 09.05.2025}
\footnotetext[9]{ \url{https://openai.com/index/gpt-4o-mini-advancing-cost-efficient-intelligence/} - Accessed 09.05.2025}

Every sub‐task in the pipeline (metadata extraction, workflow skeleton, step attributes, variable enumeration) uses a carefully crafted prompt template, combining Persona, Template, Chain-of-Thought, and Direct Knowledge Injection patterns.  By isolating the variable placeholders (playbook JSON, schema snippets, prior context fragments) from the static instruction text, our design makes it straightforward to extend or adapt the prompts for additional CACAO fields or alternative LLMs without rewriting the core template logic.  


\textbf{Experimental Setup:} We evaluate our transformation pipeline on a corpus of 40 structured SOAR playbooks from three leading vendors \cite{awesome_playbooks}: 20 Phantom playbooks, 10 Fortinet playbooks, and 10 Demisto playbooks. Each source playbook was converted from its native JSON or YAML format into a “non‐CACAO” baseline. From those dataset, we selected 10 playbooks and manually crafted CACAO translations used as ground truth for semantic evaluation. This dataset balances format diversity with a manageable size for thorough error analysis. We measure three primary metrics: the number of syntactic errors flagged by our patched CACAO JSON‐schema checker; the Damerau–Levenshtein similarity between generated and reference string fields; and the normalized Graph Edit Distance (GED) between the extracted and target workflow graphs. To capture real‐world cost, we log total API usage and compute per‐playbook expenditure (e.g.\ \$0.23 on average with GPT-4o-mini in the “all‐patterns” configuration).  Experiments involving the open source Llama3.1 model are run locally on an Intel Xeon (32 cores, 256 GB RAM) with an NVIDIA A100 GPU; all OpenAI API calls are executed from the same Linux server. As GPT inference hardware is opaque, the comparison focuses on cost and translation accuracy, not GPU seconds.

\subsection{Syntactic Accuracy}
Syntactic accuracy measures how many errors the Syntax Checker identifies in the generated translations, relying on JSON schemas defined by the OASIS Technical Committee. During implementation, however, certain discrepancies between the original schemas and CACAO specification were identified. Therefore, we created a customized version of these schemas to more precisely align with the CACAO standard and to prevent irrelevant error reporting for untranslated fields.
We use the following formula for the average syntax error. where \( d \) denotes a dataset of playbooks, \( p \) represents a single playbook, \(\text{syntax\_err\_count}(i)\) is the number of syntactic errors per playbook \( p \), and \( l(d) \) is the total number of playbooks in dataset \( d \). These metrics respectively capture the average syntactic accuracy across an entire dataset for a given model \( m \).
\[
\text{avg\_syn\_err}(m) = \frac{\sum_{p \in d}\text{syntax\_err\_count}(p)}{l(d)}
\]

\begin{figure}[ht!]
\centering
\includegraphics[width=0.6\linewidth]{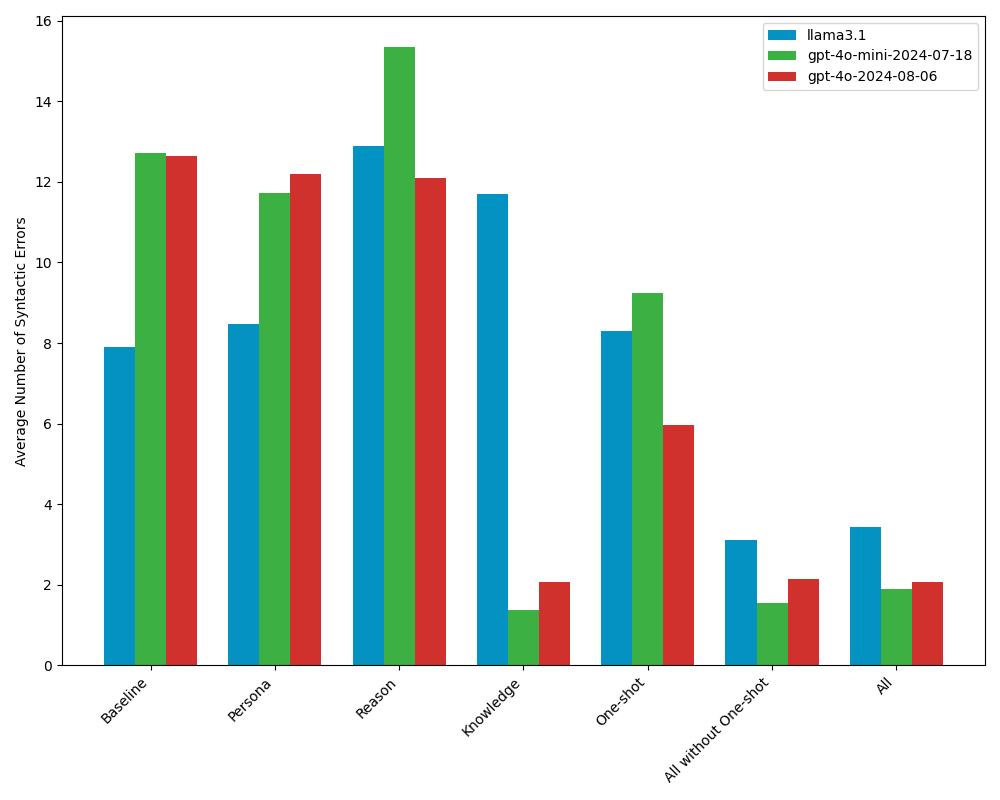}
\caption{Average syntactic errors per playbook for GPT-4o, 4o-mini, and Llama3.1-8B across different Prompt Engineering configurations (baseline, persona, reason, knowledge, one-shot, all without one-shot, and all). 
}
\label{fig:syntax_accuracy}
\end{figure}

Evaluation results indicate that using all Prompt Engineering techniques collectively reduces syntactic errors significantly, averaging a 73\% reduction across models. Figure \ref{fig:syntax_accuracy} shows the results for GPT-4o, 4o-mini, and Llama3.1-8B models. Generally, employing all Prompt Engineering techniques and all without on-shot, leads to the lowest syntactic error rate, demonstrating significant improvements in transformation accuracy. Individually, Llama3.1 initially outperformed OpenAI models in the baseline scenario, yet the inclusion of individual techniques (e.g., Persona, Reasoning) had mixed results, with Reasoning increasing errors in smaller models. Direct Knowledge Injection proved particularly effective for OpenAI models, reducing errors by approximately 83\%, likely due to their larger scale and richer pre-training data. Interestingly, adding the One-Shot example did not substantially affect overall accuracy. In general, combining multiple Prompt Engineering techniques consistently improves syntactic accuracy, especially noticeable in smaller models like Llama3.1. 
Removing the Task Decomposition step (i.e.\ issuing a single “translate whole playbook” prompt) increases the average syntax error count by a factor of three.  This ablation underscores the importance of breaking the task into focused subtasks to stay within context limits and reduce cascading errors.

\subsection{Semantic Fidelity}
The Syntax Checker validates only syntactic compliance with the CACAO standard, overlooking semantic correctness, specifically, whether the translated playbook accurately represents the logic and metadata of the original. Thus, we conducted an additional semantic evaluation to assess the fidelity of the translated playbooks to their original workflows and metadata. Semantic fidelity captures whether critical steps or properties are accurately preserved, recognizing that a syntactically valid transformation might still omit crucial details, such as necessary workflow actions. Furthermore, we carefully managed the inferential capabilities of the LLM during manual translation, penalizing unwarranted inferences to avoid inaccuracies, but permitting reasonable inference when required by the CACAO specification, for example in the case of \textit{ playbook\_types}. Notably, improved syntactic accuracy does not always correlate with semantic fidelity, forcing syntactic completeness could prompt LLMs to hallucinate information instead of correctly assigning null values. To measure semantic accuracy, we employ specific metrics that reflect these considerations.

\textbf{String-based Similarity:} We used the Damerau-Levenshtein distance~\cite{damerau_levenshtein} to quantify the differences between the manually translated and LLM extracted CACAO string fields. Because some fields employ fixed CACAO vocabularies that require exact matches (e.g., \textit{playbook\_types} and \textit{playbook\_activities}), we utilize the recall metric for those. More details on the metrics, calculations, and results are presented in the Appendix \ref{App:DL}.

\textbf{Workflow Similarity:} We employed the Graph Edit Distance (GED)~\cite{graph_edit_distance} metric to evaluate how well the workflows extracted by the LLM matched the manually translated workflows. GED measures the minimum edit cost—through node or edge insertion, deletion, or substitution—required to transform one graph into another. Each workflow is represented as an attributed relational graph, with nodes corresponding to individual workflow steps and edges representing the labeled CACAO connections (\textit{on\_completion}, \textit{on\_success}, \textit{on\_failure}, \textit{on\_true}, \textit{on\_false}). To facilitate interpretation, we normalized the GED scores using min-max normalization:

\[
\text{normalized\_GED} = \frac{x - min}{max - min}
\]

where \( x \) denotes the GED computed by the \emph{graph\_edit\_distance} function from the \textit{networkx} package\footnote{\url{https://pypi.org/project/networkx/} - Accessed 09.05.2025}, \( min = 0 \) (representing identical graphs), and \( max \) is the sum of all edges and nodes from both graphs. This normalization ensures a clear and intuitive measure of similarity between the original and translated workflows.

\begin{figure}[ht!]
\centering
\includegraphics[width=0.56\linewidth]{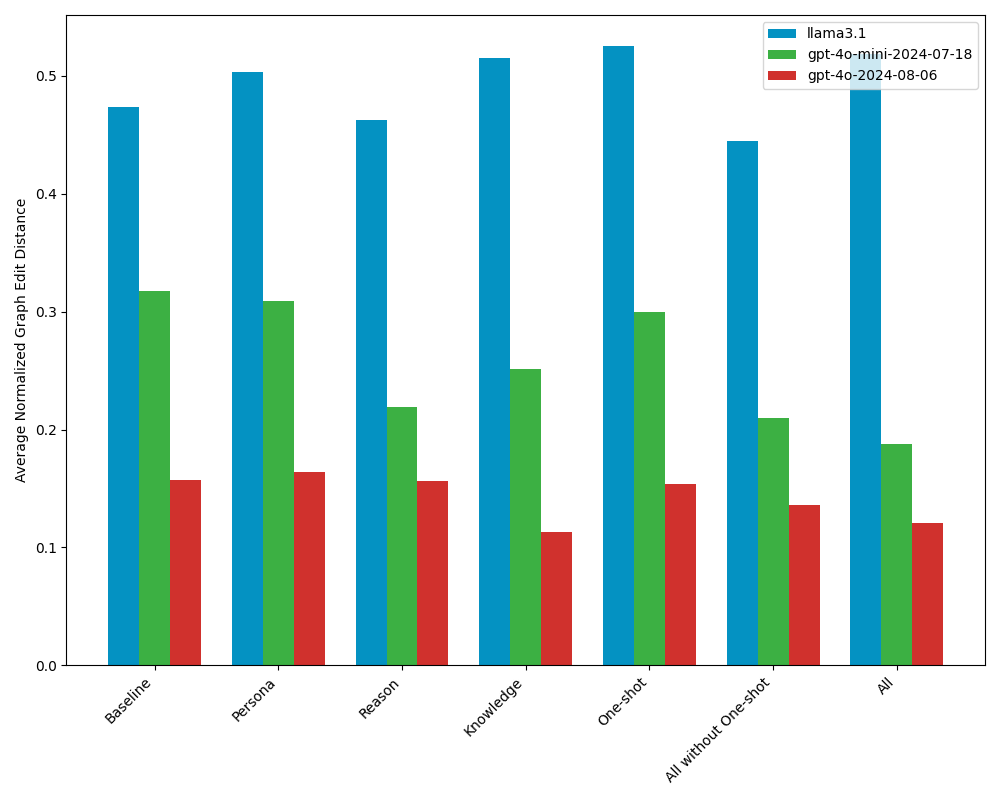}
\caption{
Average normalized Graph Edit Distance across different Prompt Engineering configurations (baseline, persona, reason, knowledge, one-shot, all without one-shot, and all). Lower distance means higher similarity.
}
\label{fig:graph_edit_distance_results}
\end{figure}

Figure \ref{fig:graph_edit_distance_results} highlights substantial differences among LLMs in accurately capturing complex workflow structures. GPT-4o achieved the best performance with the lowest average normalized Graph Edit Distance (GED) of around 0.15, indicating superior capability in accurately extracting control flows. GPT-4o-mini ranked second, exhibiting the most noticeable variation in performance when different Prompt Engineering techniques were applied, particularly with a 10\% improvement in the reasoning case. Llama3.1 consistently showed the lowest performance across all conditions. While the persona and one-shot approaches had negligible effects, the reasoning method demonstrated modest improvement, notably for GPT-4o-mini. Combining all Prompt Engineering techniques consistently yielded the greatest enhancement. Overall, these results affirm that GPT-4o effectively captures essential metadata and workflow properties with minimal benefit from additional Prompt Engineering techniques, though GPT-4o-mini clearly benefits from them. The accuracy for playbook variables was lower but deemed less critical than workflow accuracy, which GPT-4o reliably captured. Thus, GPT-4o emerges as particularly effective at maintaining semantic fidelity in translating playbook workflows.

\subsection{Syntactic Refinement}

To investigate whether the LLM could iteratively correct syntactic errors from its initial transformation, we designed an iterative refinement procedure. In this approach, we repeatedly provided the LLM with the original playbook, its previously generated transformation, detailed syntactic error messages, and selected prompt patterns (template, persona, zero-shot CoT). By applying this iterative correction process up to five times, we aimed to significantly reduce syntactic errors in the generated transformations. However, we recognize that this iterative refinement could potentially reduce semantic fidelity, as the LLM might introduce syntactically valid yet incorrect or semantically unsupported values. For simplicity and practical feasibility, we opted not to incorporate Task Decomposition into this refinement process.

\begin{figure}[ht!]
    \centering    \includegraphics[width=0.65\linewidth]{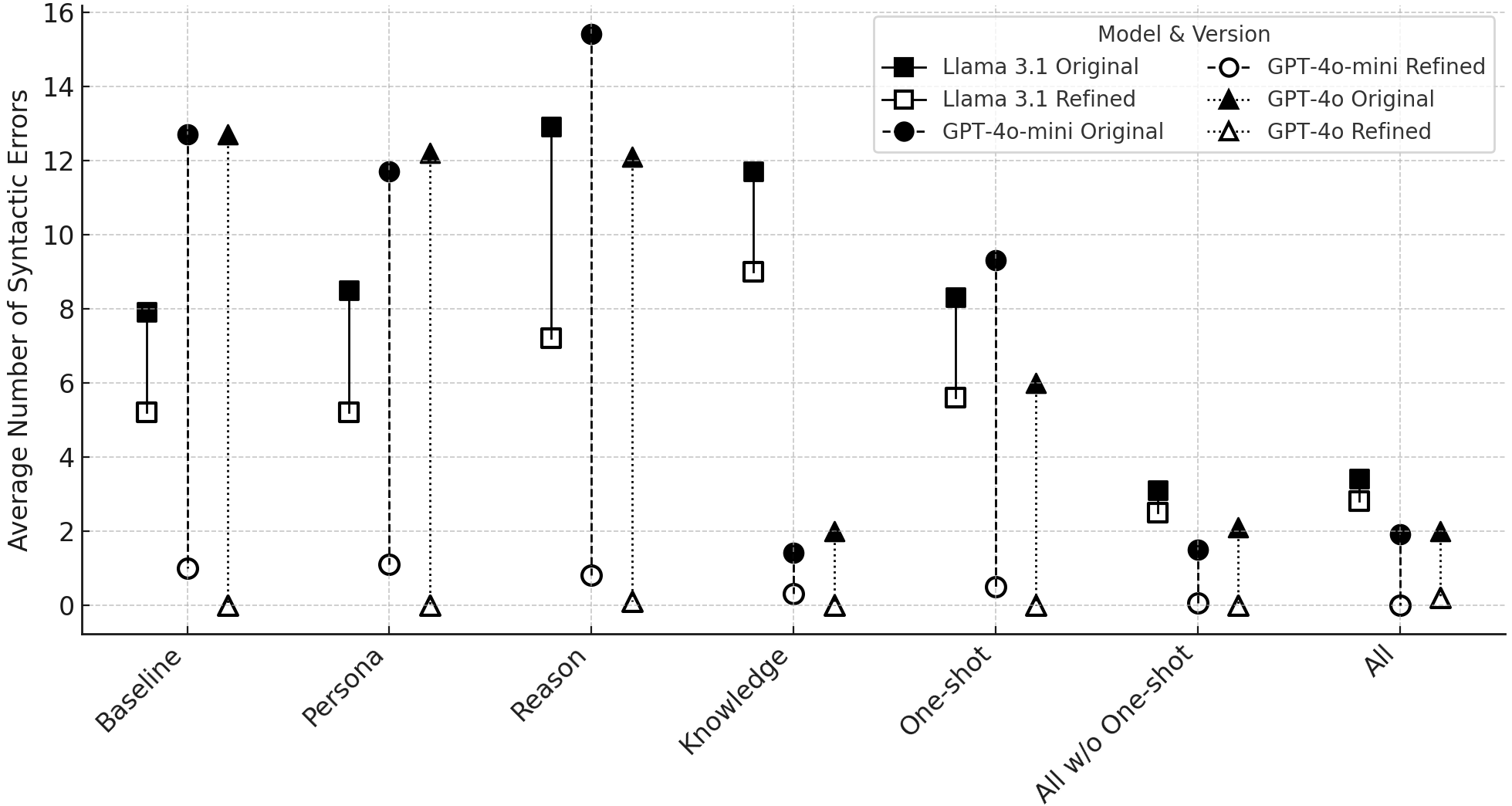}
    \caption{Comparison of average syntactic errors before and after syntactic refinement. GPT-4o and GPT-4o-mini effectively reduce errors to nearly zero after five iterations; Llama3.1 improves significantly but less effectively.}
    \label{fig:syntactic_refinement_results}
\end{figure}

The results of the syntactic refinement process, shown in Figure \ref{fig:syntactic_refinement_results}, demonstrate significant reductions in syntactic errors after iterative corrections. GPT-4o and GPT-4o-mini were highly effective, nearly eliminating syntactic errors after five iterations. Although Llama3.1 also improved notably, its performance did not match the larger OpenAI models, likely due to its smaller size and lower pre-trained knowledge capacity. Increasing iterations might further enhance Llama3.1's performance, but intrinsic limitations should be acknowledged. Appendix \ref{App:semantic-syntax} shows that iterative syntactic refinement causes negligible harm to control-flow integrity and incurs only minor semantic losses.


\section{Discussion}
\label{sec:discussion}

Our experiments highlight that \emph{Direct Knowledge Injection} is the most effective Prompt Engineering technique for reducing syntactic errors. By embedding precise snippets from the official CACAO schema directly into prompts, the LLM is effectively constrained to produce strictly schema-compliant outputs. In contrast, more generalized techniques such as \emph{naïve Retrieval-Augmented Generation (RAG)} often introduce extraneous or outdated information, potentially confusing the model and inadvertently causing hallucinations or inaccuracies \cite{rag_survey}. Direct Knowledge Injection addresses these shortcomings by providing minimal, verified context explicitly tailored to each subtask, thus ensuring a high level of syntactic accuracy.

The process of converting legacy playbooks into CACAO can be operationalized effectively through a straightforward three-step procedure. First, the legacy playbooks in non-CACAO formats (JSON, YAML, or semi-structured text) are translated using the LLM-driven pipeline to generate initial CACAO fragments. Second, a JSON schema checker validates these fragments, and identified errors or inconsistencies can be routed into a human-in-the-loop feedback mechanism for rapid correction. Finally, validated playbooks are automatically registered and integrated into the SOAR platform catalog, becoming immediately executable. This streamlined \emph{translate $\rightarrow$ validate $\rightarrow$ push} workflow facilitates efficient, near-zero-touch onboarding, potentially enabling organizations to rapidly convert and deploy hundreds of playbooks each month.

In addition to accuracy improvements, employing Prompt Engineering offers significant practical advantages over alternative approaches such as fine-tuning or embedding explicit reasoning steps directly into model training. Prompt Engineering primarily involves careful design of input prompts without modifying the underlying model, leading to substantially lower computational costs and reduced resource usage, including fewer tokens consumed during inference. This efficiency contrasts sharply with embedded reasoning methods in the state-of-the-art LLMs, higher token budgets, and considerable computational resources. It also has advantages over fine-tuning which typically require extensive datasets and longer training cycles. Thus, from an operational perspective, prompt engineered solutions provide a more lightweight, flexible, and economically viable pathway for transforming cybersecurity playbooks into standardized formats like CACAO.

Additionally, the use of cloud-based LLM APIs raises important ethical and confidentiality concerns. Incident Response playbooks often contain highly sensitive information, such as proprietary detection logic or details of customer environments, making them unsuitable for processing by third-party cloud services. To mitigate such risks, a hybrid deployment strategy is recommended. Organizations can leverage powerful cloud-based LLMs for non-sensitive transformation tasks, while reserving sensitive, classified, or regulated content for on-premise, open-source models such as Llama3.1. Supplementary safeguards, including strict logging policies, encryption, and robust access control mechanisms, further ensure the protection and integrity of playbook data throughout the transformation and deployment processes.


\textbf{Limitations:} While our results are promising, there are some limitations to consider.
First, our evaluation corpus comprises only 40 fully structured playbooks from three vendors including 10 manual CACAO translations as ground truths. This small, vendor‐specific dataset may not capture the full diversity of real‐world playbooks, particularly semi‐structured or free‐form documents frequently found in practice. Therefore, our findings may not generalize without additional data and evaluation on unstructured sources. Second, to streamline syntactic validation we patched the official CACAO 2.0 JSON schemas to ignore fields we did not extract (e.g.\ agents, targets, commands).  Although this allowed us to focus on metadata, workflow, and variables, it also means our pipeline does not yet cover the full CACAO specification.  Future work must address the complete schema and handle nested or optional entities to ensure truly end-to-end playbook transformation.


\textbf{Future Work:} An immediately actionable enhancement is to extend our pipeline to cover the remaining core CACAO entities (e.g., Agents, Targets, and Commands), which are essential for fully automated execution. Incorporating these elements would enable the end‐to‐end generation of runnable playbooks, eliminating the current need for manual post‐processing. A more ambitious but impactful direction is to embed an active learning feedback loop with SOC analysts.  By surfacing low‐confidence mappings or semantic ambiguities for rapid human correction, the system could iteratively refine its prompt strategies and fine‐tune model behavior over time, leading to continuous improvements in both syntactic and semantic accuracy.

Our current study focuses on prompt engineering because it offers a favorable cost/performance ratio, yet this does not rule out other strategies. We will explore complementary techniques, such as parameter-efficient fine-tuning, reasoning-centric prompting, and agent-based pipelines to assess when they may outperform prompt-only methods. Finally, to validate generalizability, it is critical to evaluate the approach in semi-structured or free-form playbooks, such as mixing prose, tables, and bullet lists, and to benchmark emerging LLMs such as GPT-o3, Claude 3.7, and Gemini 2.5 Pro.  Exploring how these models handle unstructured inputs and larger context windows will guide future adaptations of our transformation framework to next‐generation LLM capabilities.


\section{Conclusion}

In this work, we explored the potential of automating the transformation of legacy unstructured cybersecurity playbooks into the standardized machine-readable playbook format CACAO. We evaluated the effectiveness of our LLM-based translation process using different prompt engineering techniques, assessing the performance not only on the syntactic level but also on the semantic level.
Our results demonstrate the potential of this approach, exhibiting near-zero syntax error in our experiments. Especially the use of a capable model such as GPT-4o in combination with comprehensive prompt engineering provided promising results, reducing the syntax errors by 84\% and capturing the control-flow with a normalized graph edit distance of 0.15. In terms of CACAO's semantic fidelity, we observed the best results with the combination of Direct Knowledge Injection and Task Decomposition. Finally, we hope that our work will support practitioners interested in adopting CACAO in their security operations, serving as a step toward implementing near-zero-touch onboarding of cybersecurity playbooks into respective SOAR pipelines.

\begin{credits}
\subsubsection{\ackname} This study was Funded by the Fraunhofer Cluster of Excellence Cognitive Internet Technologies under the project name CyberGuard++.
\end{credits}

%
%
%
\bibliographystyle{splncs04}
\bibliography{ref}

\begin{thebibliography}{10}
\providecommand{\url}[1]{\texttt{#1}}
\providecommand{\urlprefix}{URL }
\providecommand{\doi}[1]{https://doi.org/#1}

\bibitem{gpt4_technical_report}
Achiam, J., Adler, S., Agarwal, S., Ahmad, L., Akkaya, I., Aleman, F.L.,
  Almeida, D., Altenschmidt, J., Altman, S., Anadkat, S., et~al.: Gpt-4
  technical report (2023)

\bibitem{akbari2022sasp}
Akbari~Gurabi, M., Mandal, A., Popanda, J., Rapp, R., Decker, S.: Sasp: a
  semantic web-based approach for management of sharable cybersecurity
  playbooks. In: Proceedings of the 17th International Conference on
  Availability, Reliability and Security. pp.~1--8 (2022)

\bibitem{llms_are_few_shot_learners}
Brown, T.B., Mann, B., Ryder, N., Subbiah, M., Kaplan, J., Dhariwal, P.,
  Neelakantan, A., Shyam, P., Sastry, G., Askell, A., Agarwal, S.,
  Herbert-Voss, A., Krueger, G., Henighan, T., Child, R., Ramesh, A., Ziegler,
  D.M., Wu, J., Winter, C., Hesse, C., Chen, M., Sigler, E., Litwin, M., Gray,
  S., Chess, B., Clark, J., Berner, C., McCandlish, S., Radford, A., Sutskever,
  I., Amodei, D.: Language models are few-shot learners (2020)

\bibitem{template_based_named_entity_recognition}
Cui, L., Wu, Y., Liu, J., Yang, S., Zhang, Y.: Template-based named entity
  recognition using bart (2021), \url{https://arxiv.org/abs/2106.01760}

\bibitem{damerau_levenshtein}
Damerau, F.J.: A technique for computer detection and correction of spelling
  errors. Commun. ACM  \textbf{7}(3),  171–176 (mar 1964).
  \doi{10.1145/363958.363994}, \url{https://doi.org/10.1145/363958.363994}

\bibitem{rag_survey}
Gao, Y., Xiong, Y., Gao, X., Jia, K., Pan, J., Bi, Y., Dai, Y., Sun, J., Wang,
  M., Wang, H.: Retrieval-augmented generation for large language models: A
  survey (2024), \url{https://arxiv.org/abs/2312.10997}

\bibitem{llama3}
Grattafiori, A., Dubey, A., Jauhri, A., Pandey, A., Kadian, A., Al-Dahle, A.,
  Letman, A., Mathur, A., Schelten, A., Vaughan, A., et~al.: The llama 3 herd
  of models (2024)

\bibitem{req}
Gurabi, M.A., Nitz, L., Bregar, A., Popanda, J., Siemers, C., Matzutt, R.,
  Mandal, A.: Requirements for playbook-assisted cyber incident response,
  reporting and automation. Digital Threats  \textbf{5}(3) (Oct 2024).
  \doi{10.1145/3688810}, \url{https://doi.org/10.1145/3688810}

\bibitem{ERCIM139}
Gurabi, M.A., Nitz, L., Joglekar, C.M., Mandal, A.: Strengthening cyber defence
  through cooperative development and shared expertise in incident response
  playbooks. ERCIM NEWS  \textbf{139},  pp. 44--46 (2024)

\bibitem{quantifying_persona_effect}
Hu, T., Collier, N.: Quantifying the persona effect in llm simulations (2024)

\bibitem{incident_response_ibm}
IBM: What is incident response?
  \url{https://www.ibm.com/topics/incident-response} (2024), accessed:
  2024-04-23

\bibitem{cacaov2_spec}
Jordan, B., Thomson, A.: Cacao security playbooks version 2.0. Latest version:
  \url{https://docs.oasis-open.org/cacao/security-playbooks/v2.0/security-playbooks-v2.0.html}
  (November 2023),
  \url{https://docs.oasis-open.org/cacao/security-playbooks/v2.0/cs01/security-playbooks-v2.0-cs01.html},
  oASIS Committee Specification 01

\bibitem{zero_shot_chain_of_thought}
Kojima, T., Gu, S.S., Reid, M., Matsuo, Y., Iwasawa, Y.: Large language models
  are zero-shot reasoners (2023)

\bibitem{process_modeling_with_llms}
Kourani, H., Berti, A., Schuster, D., van~der Aalst, W.M.P.: Process modeling
  with large language models (2024)

\bibitem{extending_context_window}
Li, R., Xu, J., Cao, Z., Zheng, H.T., Kim, H.G.: Extending context window in
  large language models with segmented base adjustment for rotary position
  embeddings. Applied Sciences  \textbf{14}(7), ~3076 (2024).
  \doi{10.3390/app14073076}, \url{https://doi.org/10.3390/app14073076}

\bibitem{method_extracting_bpmn}
Licardo, J.T.: A method for extracting bpmn models from textual descriptions
  using natural language processing. Undergraduate thesis, University of Pula
  (2023), \url{https://urn.nsk.hr/urn:nbn:hr:137:105792}

\bibitem{cybersecurity_playbook_sharing_with_stix}
Mavroeidis, V., Zych, M.: Cybersecurity playbook sharing with stix 2.1 (2022)

\bibitem{comprehensive_overview_of_llms}
Naveed, H., Khan, A.U., Qiu, S., Saqib, M., Anwar, S., Usman, M., Akhtar, N.,
  Barnes, N., Mian, A.: A comprehensive overview of large language models
  (2024)

\bibitem{SAPPAN}
Nitz, L., Akbari~Gurabi, M., Cermak, M., Zadnik, M., Karpuk, D., Drichel, A.,
  Sch\"{a}fer, S., Holmes, B., Mandal, A.: On collaboration and automation in
  the context of threat detection and response with privacy-preserving
  features. Digital Threats  \textbf{6}(1) (Feb 2025). \doi{10.1145/3707651},
  \url{https://doi.org/10.1145/3707651}

\bibitem{fine_tuning_or_retrieval}
Ovadia, O., Brief, M., Mishaeli, M., Elisha, O.: Fine-tuning or retrieval?
  comparing knowledge injection in llms (2024)

\bibitem{temperature_creativity}
Peeperkorn, M., Kouwenhoven, T., Brown, D., Jordanous, A.: Is temperature the
  creativity parameter of large language models? (2024),
  \url{https://arxiv.org/abs/2405.00492}

\bibitem{extracting_accurate_materials_data}
Polak, M.P., Morgan, D.: Extracting accurate materials data from research
  papers with conversational language models and prompt engineering. Nature
  Communications  \textbf{15}(1) (Feb 2024). \doi{10.1038/s41467-024-45914-8},
  \url{http://dx.doi.org/10.1038/s41467-024-45914-8}

\bibitem{systematic_survey_prompt_engineering_in_llms}
Sahoo, P., Singh, A.K., Saha, S., Jain, V., Mondal, S., Chadha, A.: A
  systematic survey of prompt engineering in large language models: Techniques
  and applications (2024)

\bibitem{graph_edit_distance}
Sanfeliu, A., Fu, K.S.: A distance measure between attributed relational graphs
  for pattern recognition. IEEE Transactions on Systems, Man, and Cybernetics
  \textbf{SMC-13}(3),  353--362 (1983). \doi{10.1109/TSMC.1983.6313167}

\bibitem{awesome_playbooks}
Schlette, D., Empl, P., Caselli, M., Schreck, T., Pernul, G.: Do you play it by
  the books? a study on incident response playbooks and influencing factors.
  In: Proceedings of the 45th IEEE Symposium on Security and Privacy, {SP}
  2024, San Francisco, CA, USA, May 20-23, 2024. pp. 1--19. {IEEE} (2024)

\bibitem{tsirakis2025}
Tsirakis, O., Fysarakis, K., Mavroeidis, V., Papaefstathiou, I.:
  Operationalizing cybersecurity knowledge: Design, implementation \&
  evaluation of a knowledge management system for cacao playbooks.
  \url{https://arxiv.org/pdf/2503.05206} (2025), arXiv preprint
  arXiv:2503.05206

\bibitem{attention_is_all_you_need}
Vaswani, A., Shazeer, N., Parmar, N., Uszkoreit, J., Jones, L., Gomez, A.N.,
  Kaiser, L., Polosukhin, I.: Attention is all you need (2023)

\bibitem{prompt_engineering_approach_structured_data_extraction}
Vijayan, A.: A prompt engineering approach for structured data extraction from
  unstructured text using conversational llms. In: Proceedings of the 2023 6th
  International Conference on Algorithms, Computing and Artificial
  Intelligence. p. 183–189. ACAI '23, Association for Computing Machinery,
  New York, NY, USA (2024). \doi{10.1145/3639631.3639663},
  \url{https://doi.org/10.1145/3639631.3639663}

\bibitem{finetuned_llms_are_zero_shot_learners}
Wei, J., Bosma, M., Zhao, V.Y., Guu, K., Yu, A.W., Lester, B., Du, N., Dai,
  A.M., Le, Q.V.: Finetuned language models are zero-shot learners (2022)

\bibitem{chain_of_thought}
Wei, J., Wang, X., Schuurmans, D., Bosma, M., Ichter, B., Xia, F., Chi, E., Le,
  Q., Zhou, D.: Chain-of-thought prompting elicits reasoning in large language
  models (2023)

\bibitem{prompt_pattern_catalog}
White, J., Fu, Q., Hays, S., Sandborn, M., Olea, C., Gilbert, H., Elnashar, A.,
  Spencer-Smith, J., Schmidt, D.C.: A prompt pattern catalog to enhance prompt
  engineering with chatgpt (2023)

\bibitem{least_to_most_prompting}
Zhou, D., Schärli, N., Hou, L., Wei, J., Scales, N., Wang, X., Schuurmans, D.,
  Cui, C., Bousquet, O., Le, Q., Chi, E.: Least-to-most prompting enables
  complex reasoning in large language models (2023)

\end{thebibliography}

\appendix 

\section{String-based Similarity Results} 
\label{App:DL}

Damerau-Levenshtein distance was used to quantify differences between manually translated and LLM-extracted CACAO string fields. This metric measures the minimum number of operations (insertion, deletion, substitution, transposition) required to convert one string into another. Given two strings \(a\) and \(b\), we define the Damerau-Levenshtein similarity metric \( S(a, b) \) as follows.
\[
 S(a, b) =
    \begin{cases}
        0 & \text{if either } a \text{ or } b \text{ are empty strings}, \\
        1 - \frac{D(a, b)}{\max(\lvert a \rvert, \lvert b \rvert)} & \text{otherwise},
    \end{cases}
\]

where \(D(a,b)\) is the Damerau-Levenshtein distance (calculated by the jellyfish package\footnote{\url{https://pypi.org/project/jellyfish/} - Accessed 09.05.2025}) and \(\lvert a \rvert, \lvert b \rvert\) represent the lengths of the respective strings. For fields constrained by CACAO vocabularies, such as \textit{playbook\_types} and \textit{playbook\_activities}, we use recall instead, defined as:

\[
 \text{Recall}(f) = \frac{TP(f)}{TP(f) + FN(f)}
\]

where \(TP\) and \(FN\) represent True Positives and False Negatives for field \(f\). The final semantic fidelity is the average of these metrics in all relevant fields \(F\):
\[
 \text{accuracy}(F) = \frac{1}{|F|} \sum_{f \in F}\left\{
    \begin{array}{ll}
        S(f_{ground\_truth}, f_{transformation}) & \text{if } f_{type} = \text{string}, \\
        \text{Recall}(f_{ground\_truth}, f_{transformation}) & \text{otherwise},
    \end{array} \right.
\]

where \(f_{ground\_truth}\) and \(f_{transformation}\) denote the manual and generated translations respectively, and \(f_{type}\) indicates the field type. 

\begin{figure}[ht!]
    \centering
    \begin{subfigure}[b]{0.47\textwidth}
        \includegraphics[width=\textwidth]{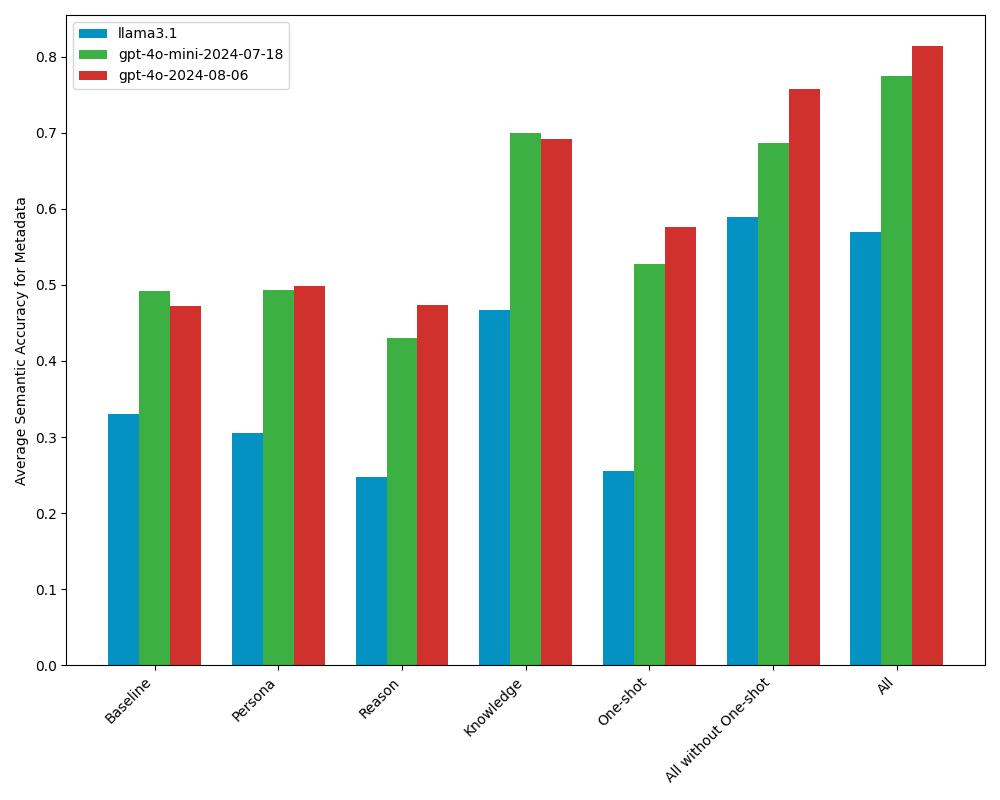}
        \caption{}
        \label{fig:metadata_results}
    \end{subfigure}
    \hfill
    \begin{subfigure}[b]{0.47\textwidth}
        \includegraphics[width=\textwidth]{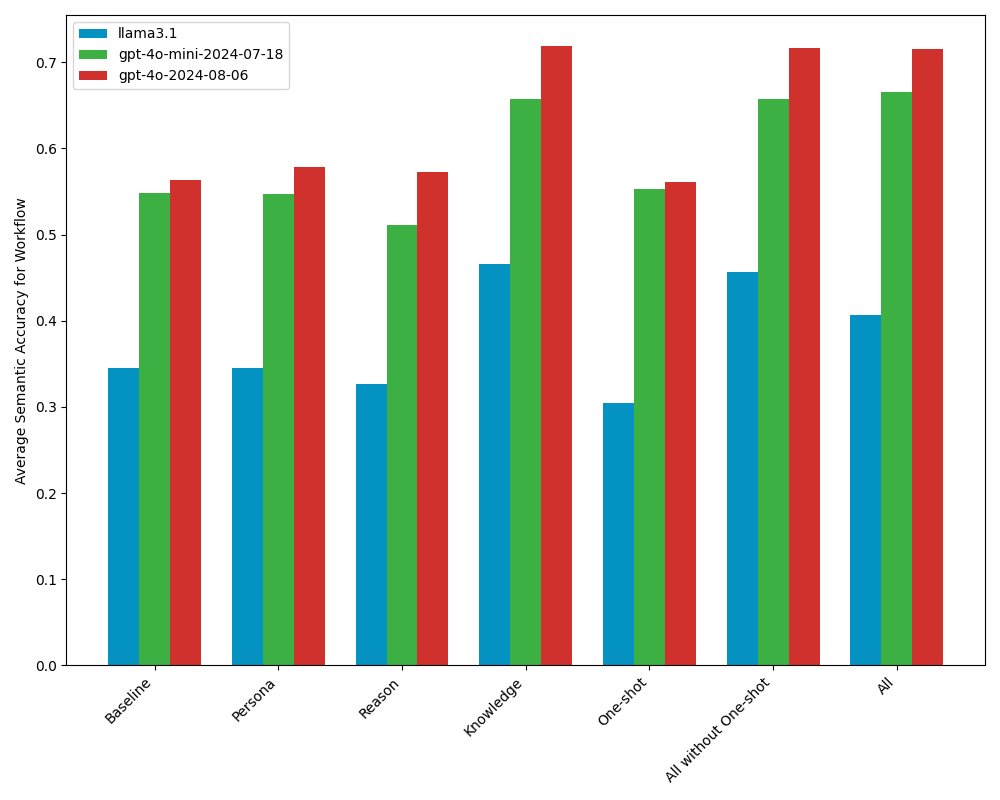}
        \caption{}
        \label{fig:workflow_results}
    \end{subfigure}

    \vskip\baselineskip

    \begin{subfigure}[b]{0.47\textwidth}
        \includegraphics[width=\textwidth]{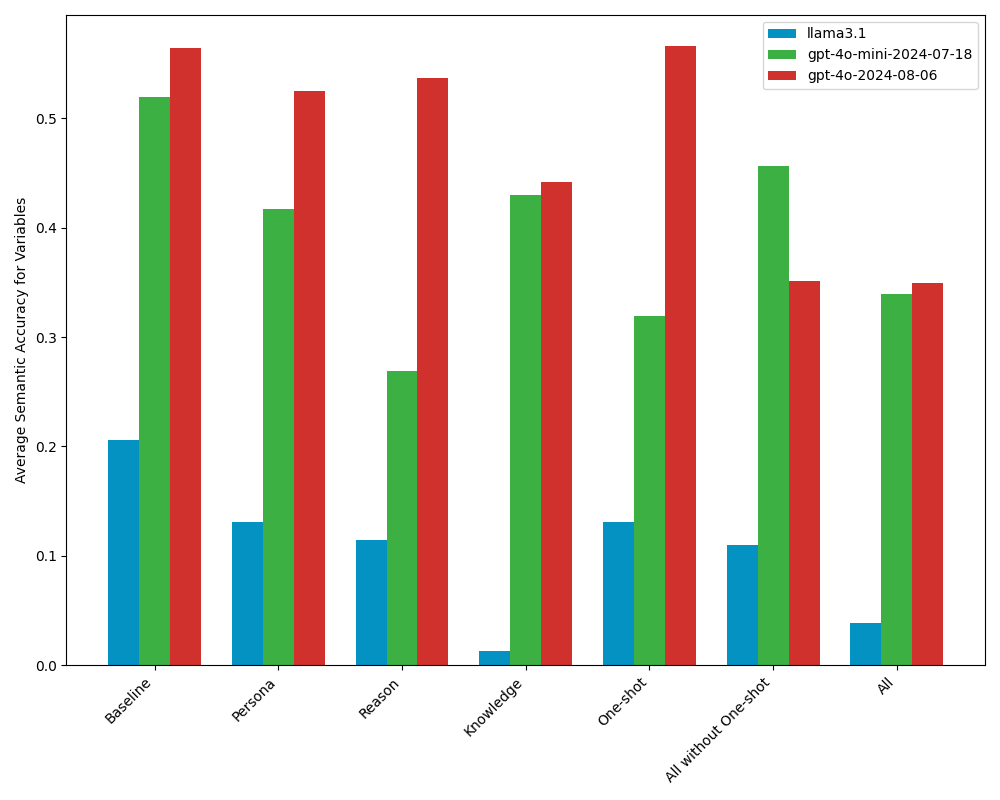}
        \caption{}
        \label{fig:variables_results}
    \end{subfigure}

    \caption{Semantic fidelity evaluation for a) metadata, b) workflow, and c) playbook variable fields across different Prompt Engineering techniques.
    }
    \label{fig:semantic_accuracy_results}
\end{figure}

Figure \ref{fig:metadata_results} presents the semantic fidelity for metadata extraction across different Prompt Engineering scenarios. Direct Knowledge Injection significantly improved accuracy by over 20\% for the OpenAI models, while adding the one-shot example to the combined techniques provided a slight additional gain. Similar to the syntactic results, individual Prompt Engineering methods, except for Direct Knowledge Injection, slightly decreased semantic fidelity for Llama3.1. Notably, the most powerful model, GPT-4o, showed a substantial improvement exceeding 30\% when all Prompt Engineering techniques were utilized together. Figure \ref{fig:workflow_results} shows that, unlike metadata extraction, adding the one-shot example did not notably enhance workflow-field accuracy when combined with other Prompt Engineering techniques. Direct Knowledge Injection again had the strongest positive effect, improving semantic fidelity across all models—most notably for GPT-4o, with an increase of up to 15\%. The Persona, Reasoning, and One-Shot techniques individually showed minimal to no impact. As anticipated, GPT-4o consistently achieved the highest semantic fidelity, followed by GPT-4o-mini. Figure \ref{fig:variables_results} illustrates a distinct trend for semantic fidelity regarding playbook variables compared to earlier evaluations. Surprisingly, the baseline case outperformed scenarios utilizing additional Prompt Engineering techniques across all models, suggesting that these methods may hinder rather than enhance the accurate extraction of variables. Direct Knowledge Injection, notably, resulted in an accuracy decline exceeding 18\% for Llama3.1. Furthermore, a substantial 22\% accuracy difference between GPT-4o and GPT-4o-mini in the reasoning and one-shot cases implies that complex reasoning tasks and illustrative examples are more effectively handled by larger models. We attribute the observed performance decrease to the LLMs' increasing difficulty in clearly distinguishing between CACAO workflow step variables and general playbook variables when more Prompt Engineering techniques are introduced.

\section{Impact of Syntactic Refinement on Semantic Fidelity}
\label{App:semantic-syntax}
\begin{figure}[ht!]
    \centering
    \begin{subfigure}[b]{0.45\textwidth}
        \centering
        \includegraphics[width=\textwidth]{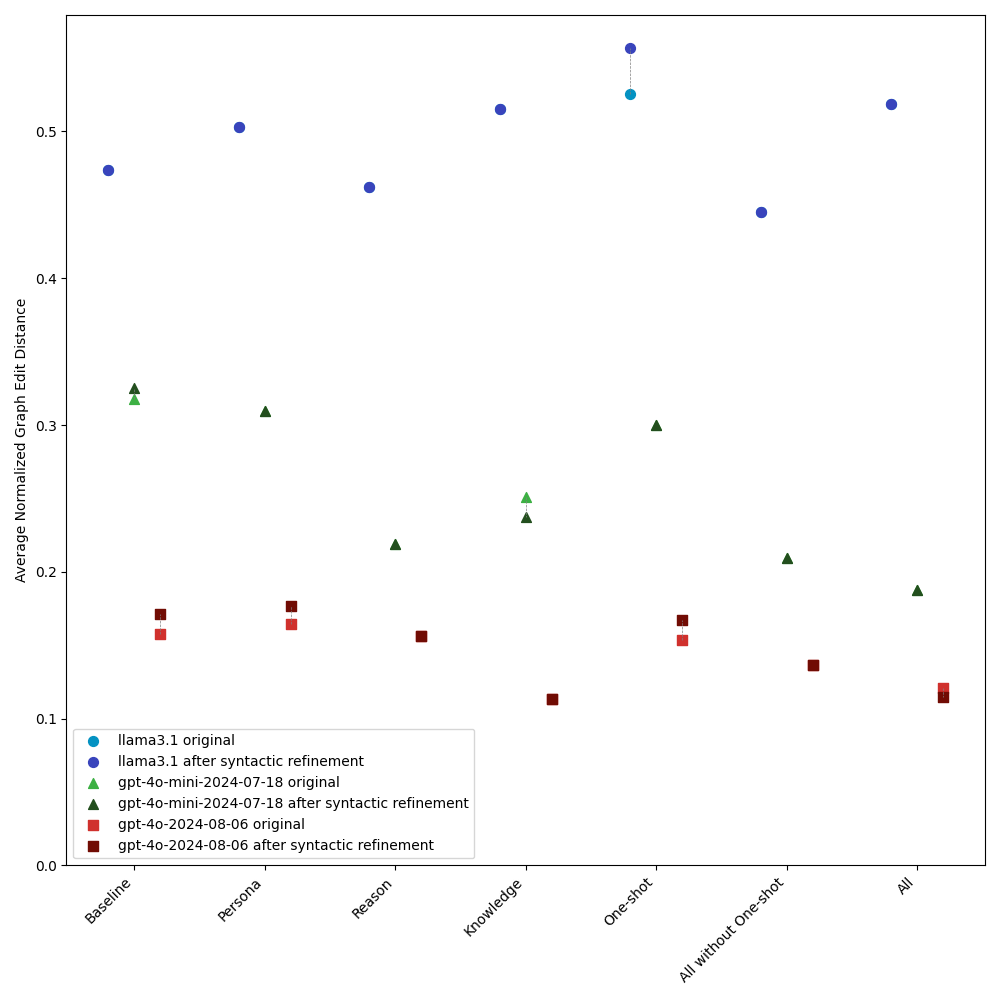}
        \caption{
        }
        \label{fig:syntactic_refinement_graph_edit_distance}
    \end{subfigure}
    \hfill
    \begin{subfigure}[b]{0.45\textwidth}
        \centering
        \includegraphics[width=\textwidth]{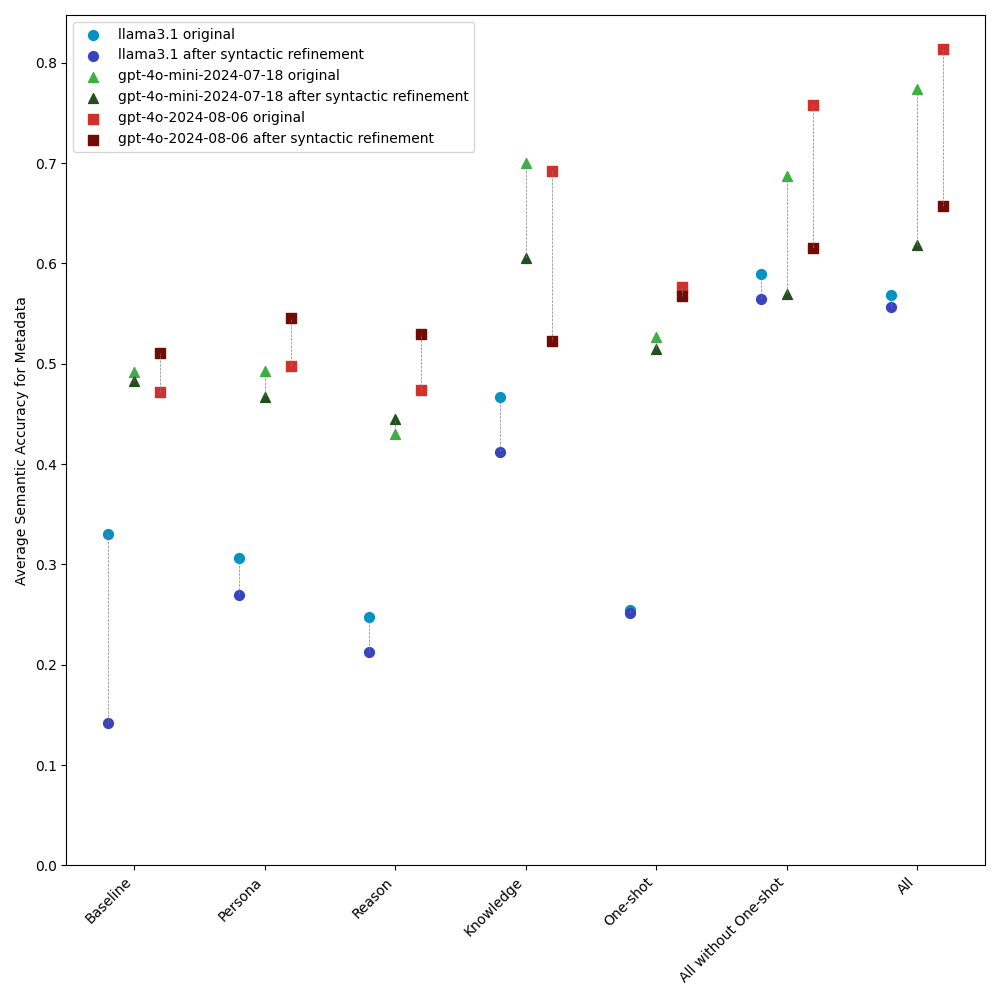}
        \caption{
        }
        \label{fig:syntactic_refinement_metadata}
    \end{subfigure}
    \vskip\baselineskip
    \begin{subfigure}[b]{0.45\textwidth}
        \centering
        \includegraphics[width=\textwidth]{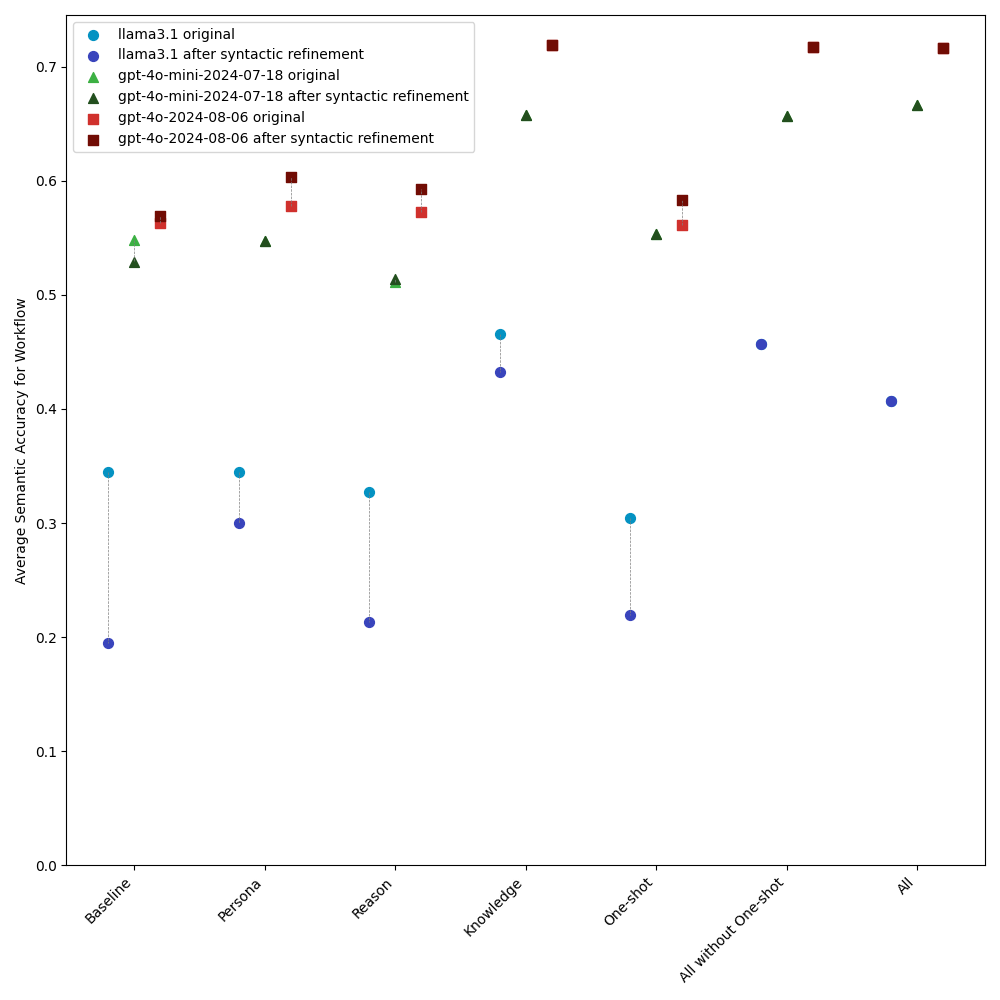}
        \caption{
        }
        \label{fig:syntactic_refinement_workflow}
    \end{subfigure}
    \hfill
    \begin{subfigure}[b]{0.45\textwidth}
        \centering
        \includegraphics[width=\textwidth]{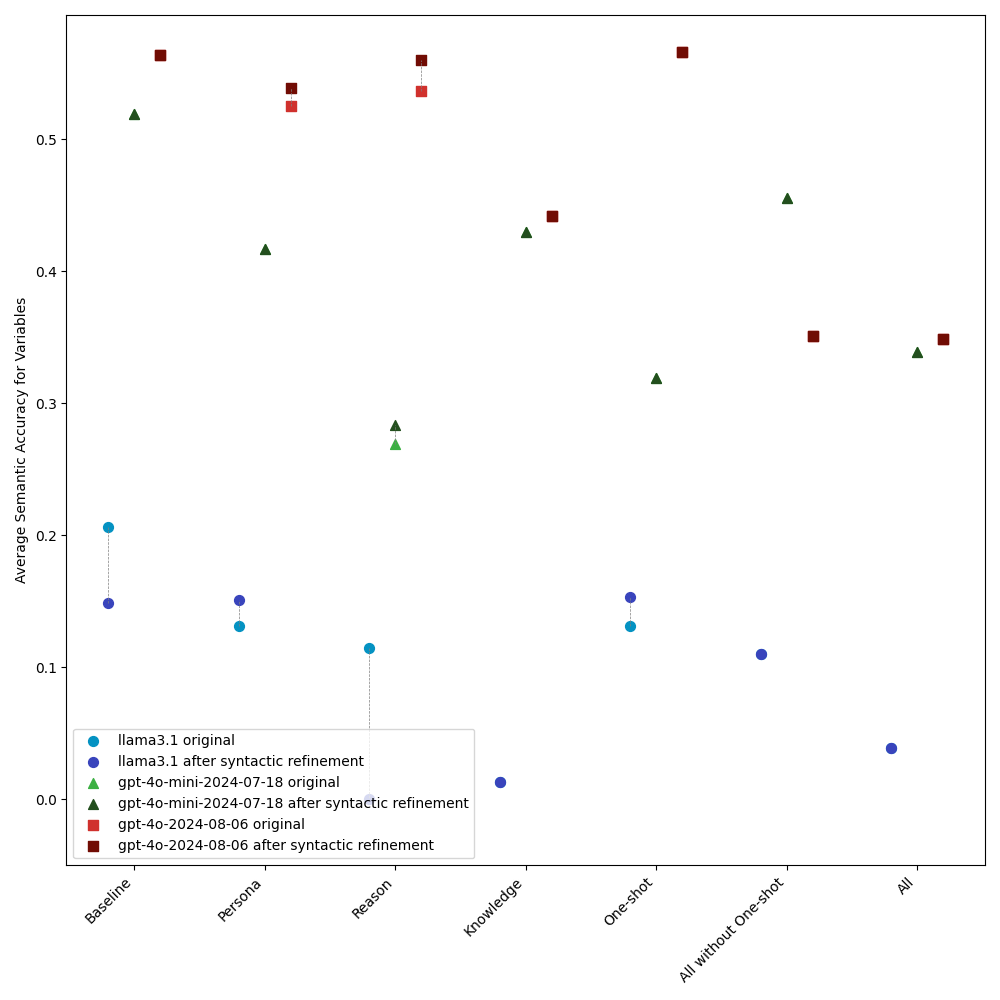}
        \caption{
        }
        \label{fig:syntactic_refinement_variables}
    \end{subfigure}
    
    \caption{Semantic impact of syntactic refinement: (a) The graph edit distance stays constant, confirming that control-flow semantics are preserved. Minor drops in semantic fidelity appear for (b) metadata, (c) workflow fields, and (d) playbook variables, more noticeably with Llama, while remaining negligible for GPT.}
    \label{fig:syntactic_refinement_semantic}
\end{figure}

Figures \ref{fig:syntactic_refinement_semantic} illustrate the semantic impact of the refinement process. It shows that syntactic refinement does not degrade the control flow (i.e., GED remains stable in (a)) while only marginally lowering metadata fidelity (b), with maximum reductions of around 20\% (Llama3.1 baseline), 19\% (GPT-4o, knowledge case), and 15\% (GPT-4o-mini, combined case). Similarly, certain cases showed moderate decreases in semantic fidelity for workflow fields (c) and playbook variables (d), especially with Llama3.1. Despite these minor trade-offs, the substantial syntactic improvements generally justify the iterative refinement process. After reviewing these results, we can confidently state that the average number of syntactic errors for OpenAI models is nearly zero across all scenarios. Llama3.1 also manages to achieve a similar reduction, although to a lesser degree.

\end{document}